\documentclass[a4paper]{article}

\usepackage{INTERSPEECH_v2}
\usepackage{graphicx}
\usepackage[table]{xcolor}
\usepackage{hhline}

\def\x{\mathbf{x}}
\def\y{\mathbf{y}}
\def\z{\mathbf{z}}
\def\h{\mathbf{h}}

\definecolor{Orange}{rgb}{1,0.75,0}

\title{Investigation of Using VAE for i-Vector Speaker Verification}
\name{Timur Pekhovsky$^{1,2}$, Maxim Korenevsky$^{1,2}$}
\address{
  $^1$STC-Innovations Ltd., St. Petersburg, Russia\\
  $^2$ITMO University, St. Petersburg, Russia}
\email{tim,korenevsky@speechpro.com}

\begin{document}

\maketitle
\begin{abstract}
New system for i-vector speaker recognition based on variational autoencoder (VAE) is investigated. VAE is a promising approach for developing accurate deep nonlinear generative models of complex data. Experiments show that VAE provides speaker embedding and can be effectively trained in an unsupervised manner. LLR estimate for VAE is developed. Experiments on NIST SRE 2010 data demonstrate its correctness. Additionally, we show that the performance of VAE-based system in the i-vectors space is close to that of the diagonal PLDA. Several interesting results are also observed in the experiments with $\beta$-VAE. In particular, we found that for $\beta\ll 1$, VAE can be trained to capture the features of complex input data distributions in an effective way, which is hard to obtain in the standard VAE ($\beta=1$).

\end{abstract}

\section{Introduction}
In recent years the promising deep generative model VAE (variational autoencoder) was developed~\cite{Kingma,Rezende} which has the following properties: 
(i) it can be made sufficiently deep to capture the complex data structures; (ii) it provides fast sampling of data from the inference model and (iii) it is computationally feasible and scalable.

This paper presents the attempt to apply this VAE model in i-vectors space~\cite{Dehak} for the speaker verification task.
We deliberately chose these features in spite of the fact that they are highly Gaussian after i-vector extractor~\cite{Dehak} and length normalization~\cite{Garcia} and are ideal for subsequent modeling with Gaussian PLDA (Probabilistic Linear Discriminant Analysis)~\cite{PLDA}. So it should be expected that for these features the performance of VAE will be limited by that of PLDA.

The main goal of this paper is to develop and assay the verification backend for VAE. 
It is convenient to solve this task in the i-vectors space and then to extend the solution to other features.

\section{Verification system based on VAE}
In this paper we confine to the investigation of the simplest diagonal version of VAE with a single hidden stochastic layer.

\subsection{VAE}
\begin{figure}[!ht]
  \centering
  \includegraphics[width=\linewidth]{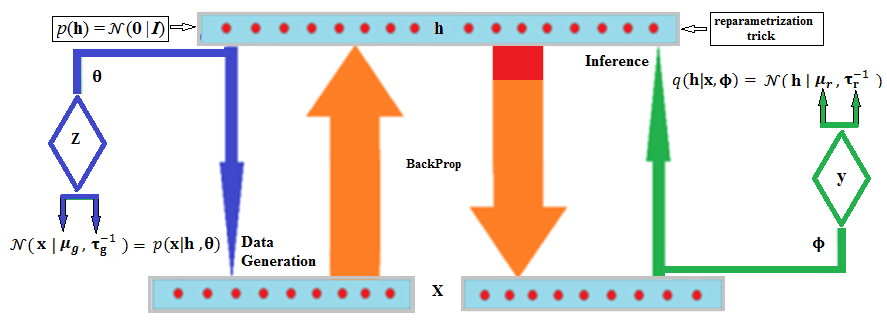}
  \caption{Schematic diagram of VAE. The red part of arrow indicates that  gradient is taken with respect to latent variables.}
  \label{fig1}
\end{figure}

It is more convenient to consider VAE as a model of deep nonlinear factor analysis (FA), though its original name~\cite{Kingma} suggests the obvious relation to conventional {autoencoders}~\cite{Autoenc}. In autoencoders all hidden layers consist of only deterministic neurons whereas for the factor analysis latent variable we need at least one hidden layer consisting of stochastic neurons (see Figure~\ref{fig1} where it is denoted as $\mathbf h$).

Similar to the classic factor analysis we should be able to perform the following actions: (i) to make inference for the latent variable posterior and (ii) to sample observed data vectors $X$. To meet these requirements VAE comprises two neural nets, namely {inference net} and {generative net} shown at the right and left parts of Figure~\ref{fig1} respectively. Both of them involve the Gaussian assumptions and have identical structure.  In addition to the the input layer $X$ of size $D_x$ and stochastic layer $\mathbf h$ of size $D_h$, this structure contains the layers $\mathbf z$ and $\mathbf y$ of deterministic neurons of size $D_d$, shown on Figure~\ref{fig1} as rhombs. These layers in VAE are responsible for the additional depth and for the nonlinearity:
\begin{equation}
   {\mathbf z} = \tanh \left[ \h{W_v^{(\theta)}} + b_v^{(\theta)} \right] =
       \tanh \left[ \widetilde{\h} \widetilde W_v^{(\theta)}\right],
\label{eq1}       
\end{equation}
\begin{equation}
   {\mathbf y} = \tanh \left[ \x {W_v^{(\phi)}} + b_v^{(\phi)} \right] =
       \tanh \left[ \widetilde {\mathbf x} \widetilde W_v^{(\phi)}  \right],
\label{eq2}       
\end{equation}
while parameters of the mean vectors and precision matrices of both generation and inference nets are computed with the linear connections only:
\begin{equation}
   \mu_g({\mathbf h},\theta) = 
                               \widetilde {\mathbf z} \widetilde W_\mu^{(\theta)} ,\quad
   \tau_g({\mathbf h},\theta) =
   \exp\left[ \widetilde {\mathbf z} \widetilde W_\mu^{(\theta)}\right],
\label{eq4}                               
\end{equation}
\begin{equation}
   \mu_r({\mathbf x},\phi) = 
           \widetilde {\mathbf y} \widetilde W_\mu^{(\phi)}, \quad
   \tau_r({\mathbf x},\phi) = 
   \exp\left[ \widetilde {\mathbf y}  \widetilde W_\mu^{(\phi)} \right],
\label{eq6}                               
\end{equation}
where indices $r$ and $g$ are used for the inference and generation nets respectively. Hereinafter all vectors are treated as \textbf{row} vectors.
In the expressions~(\ref{eq1}--\ref{eq6}) the entire set of the generated net's parameters is denoted as $\theta$ and that of the inference net is denoted as $\phi$, following the original paper~\cite{Kingma}, and the additional notations like $\widetilde \h\equiv [\h\ \ 1]$ and 
$\widetilde W_v^{(\theta)}\equiv \left[{W_v^{(\theta)}}^T\ \  {b_v^{(\theta)}}^T\right]^T$ are used.
We consider only diagonal precision matrices $\tau_g$ and $\tau_r$ also treated as vectors. This is what we mean by diagonality of VAE.


\subsection{Learning VAE}
Let $\mathbf{X}=\left\{\mathbf{x}^{(i)} \right\}_{i=1}^N$ be a training set of i-vectors of the dimension $D_x$. Due to the nonlinearity and depth it is difficult to maximize likelihood directly via analytical EM-algorithm. That is why the authors of~\cite{Kingma,Rezende} have to use the VBA approximation. In the context of analogy to factor analysis we are comparing VAE to FA-VBA.
 
Following~\cite{Bishop}, let us separate the {lower bound} $\mathcal{L}(\mathbf{x})$ from the {evidence} 
$\log p(\x)$:
\begin{equation}
   \log p(\x) = \mathcal{L}(\x) + D_{KL}[q(\h | \x) \parallel p(\h | \x)],
\label{eq7}                               
\end{equation}
where {lower bound} is
\[
   \mathcal{L}(\x)  =  \mathbb{E}_{q(\h|\x)}\left[ \log\frac{p(\x,\h)}{q(\h|\x)}\right] 
   =  \mathbb{E}_{q(\h|\x)}\left[ \log\frac{p(\x | \h)p(\h) }{q(\h|\x)}\right]
\]
\begin{equation}
                   =   \mathbb{E}_{q(\h\mid\x)}\left[ p(\x | \h) \right] 
                                - D_{KL}\left[q(\h|\x) \mid\mid p(\h)\right]
\label{eq8}                               
\end{equation}
The true posterior $p(\h|\x)$ in~(\ref{eq7}) is intractable, therefore it is approximated by the variational posterior $q(\h|\x)$.

Like in the conventional Gaussian FA-VBA, the hidden variable prior is assumed to be $p(\h)=\mathcal{N}(0|\mathbf{I})$, the posteriors $q(\h|\x,\phi)$ and $p(\x|\h(\phi), \theta)$ are Gaussian and the KL-divergence can found analytically 
and does not depend on $\h$.
As in FA-VBA we need to maximize the lower bound $\mathcal{L}(\x)$ to solve the optimization task for the VAE parameters $\Psi=\{\theta, \phi\}$. Due to nonlinearity and depth we are now unable to find this VBA-solution analytically. So we have to resort to the search for the stationary point $\Psi_0$ using the numerical stochastic gradient $\nabla_{\Psi}\mathcal{L}(\x)$ ascent to update parameters. We also should be able to sample from $q(\h|\x)$ during the inference stage.

The computation of gradient $\nabla_{\theta}\mathcal{L}(\x^{(i)})$ does not reveal any difficulties so the standard deterministic backpropagation can be used. However the gradient $\nabla_{\phi}\mathcal{L}(\x^{(i)})$ looks more problematic.
It is known that the na\"ive Monte Carlo approximation of expectation in~(\ref{eq8}) which uses $K$ samples directly from the inference net $\h^{(k)}\sim q(\h|\x^{(i)},\phi)$ results in very high variance~\cite{Blei}. 
In this case the training is slow because the gradients of 
$\log p(\x|\h(\phi),\theta)$ with respect to latent variable $\h$ are not used~\cite{Dayan, Mnih}. In the papers~\cite{Kingma,Rezende} the {reparametrization trick} was proposed, according to which the vectors $\h^{(i,k)}$ for the Monte-Carlo estimation are not sampled from $q(\h|\x^{(i)},\phi)$ but instead generated from the deterministic transform
\[
  \h^{(i,k)} = \mu_r({\mathbf x^{(i)}},\phi) + \left[\tau_r({\mathbf x^{(i)}},\phi)\right]^{-1/2} \odot \epsilon^{(k)},
\]
where $\epsilon^{(k)}$ are sampled from the fixed distribution $\epsilon^{(k)} \sim \mathcal{N}(0,\mathbf{I})$.
Using the reparametrization trick makes it possible to push gradient $\nabla_{\phi}$ inside the expectation in~(\ref{eq8}) because it is now taken over the fixed distribution of $\epsilon$ which is independent of $\phi$. As a result, the final expression for the gradient of~(\ref{eq8}) is as follows:
\[
   \nabla_{\Psi}\mathcal{L}(\x^{(i)}) \approx 
   \frac 1K \sum_{k=1}^K \left[\nabla_{\Psi} \log p(\x^{(i)}|\h^{(i,k)})\right]-
\]
\[
   -\nabla_{\Psi}D_{KL}\left[q(\h|\x^{(i)}) \mid\mid p(\h)\right],
\]
where $p(\x^{(i)}|\h^{(i,k)})$ is Gaussian with parameters $\mu_g$ and $\tau_g$.
Ultimately, we have the following expressions for the gradients of $\mathcal{L}(\x^{(i)})$ with respect to $\theta$:
\[
   \frac{\partial \mathcal{L}(\x^{(i)})}{\partial \widetilde{W}_{\mu}^{(\theta)}} = 
\mbox{$\widetilde{\z}^{(i,k)}$}^TA,\qquad
   \frac{\partial \mathcal{L}(\x^{(i)})}{\partial \widetilde{W}_{\mu}^{(\theta)}} = 
\mbox{$\widetilde{\z}^{(i,k)}$}^TB,
\]
\[
   \frac{\partial \mathcal{L}(\x^{(i)})}{\partial \widetilde{W}_{v}^{(\theta)}} = 
\mbox{$\widetilde{\h}^{(i,k)}$}^TG,
\]
and with respect to $\phi$:
\[
   \frac{\partial \mathcal{L}(\x^{(i)})}{\partial \widetilde{W}_{\mu}^{(\phi)}} =
   \mbox{$\widetilde{\y}^{(i)}$}^T  (S-\mu_r),
   \quad
   \frac{\partial \mathcal{L}(\x^{(i)})}{\partial \widetilde{W}_{\tau}^{(\phi)}} = 
   \mbox{$\widetilde{\y}^{(i)}$}^T  (S\odot F+R),
\]
\[
   \frac{\partial \mathcal{L}(\x^{(i)})}{\partial \widetilde{W}_{v}^{(\phi)}} = 
   \mbox{$\widetilde{\x}^{(i)}$}^T
     \left\{\left(
     \left[ S\odot F \right]  {W_{\tau}^{(\phi)}}^T + S{W_{\mu}^{(\phi)}}^T
     \right)\odot T+\right.
\]
\[
  +\left.\left( R  {W_{\tau}^{(\phi)}}^T - \mu_r  {W_{\mu}^{(\phi)}}^T
  \right)\odot T\right\},    
\]
where
\[
  A = (\x^{(i)}-\mu_g)\odot\tau_g, \quad
  B=\frac12\left[E_x - (x^{(i)}-\mu_g)\odot A   \right],
\]
\[
   R = \frac12\left[\tau_r^{-1}-E_h\right] , \qquad
   F = -\frac12\left[\tau_r^{-1/2}\odot \epsilon^{(k)}\right],
\]
\[
   C = 1-\tanh^2{\z}^{(i,k)}, \qquad
   T = 1-\tanh^2{\y}^{(i)},
\]
\[
   G = C \odot \left( B  {W_{\tau}^{(\theta)}}^T + A  {W_{\mu}^{(\theta)}}^T  \right), \qquad
   S = G {W_{v}^{(\theta)}}^T,
\]
\[
   E_h = [{\mathrm{diag^{-1}}(I_{D_h})}]^T, \qquad E_x = [{\mathrm{diag^{-1}}(I_{D_x})}]^T.
\]
We found that $K=1$ and minibatch size 100 provided the best results. Moreover, when we trained VAE  with $K=10$ and $K=100$ it ceased to capture a complex structure of data and was able to generate data only from a  distribution like a single Gaussian (which corresponds to the classical FA). The similar situation was observed when using a na\"ive Monte Carlo estimate of expectation in~(\ref{eq8}) instead of reparametrization trick.


\subsection{RMS-prop optimizer}
The choice of the optimizer like SGD or AdaGrad~\cite{DHS10} is crucial for training VAE. In this work we used the RMS-prop optimizer~\cite{Hinton}:
\[
   MS_j^{(new)} = \gamma   MS_j^{(old)} + 
   (1-\gamma)\left[\partial\mathcal{L}(x^{(i)})/\partial \Psi_j\right]^2,
\]
where $0\leqslant \gamma\leqslant 1$. We divide gradient with respect to the parameter $\Psi_j$ by square root of the smoothed value $MS_j^{(new)}$.


\subsection{LLR scoring for VAE}
Since VAE is a discriminative model we can only use {evidence} or {marginal likelihood} to obtain the speaker verification scores. Thus our verification score for the pair of i-vectors $\{\x_1=\x_{test}, \x_2=\x_{enroll}\}$ is a {Likelihood Ratio}:
\begin{equation}
  LR(\x_1,\x_2)=\frac{P(\x_1,\x_2| H_{tar})}{P(\x_1,\x_2| H_{imp})}=
  \frac{P(\x_1,\x_2| \theta)}{P(\x_1| \theta)  P(\x_2| \theta)},
\label{eq9}  
\end{equation}
where $H_{tar}, H_{imp}$ --- are the hypotheses about the facts that $\x_1, \x_2$
are related to the same or different speakers respectively.

If only a single latent variable $\h$ is used then we can estimate the marginal likelihood under impostor hypothesis with the help of the importance sampling which uses $q(\h|\x,\phi)$ as a proposal distribution:
\[
  P(\x|\theta) = 
  \int \frac{p(\x|\h,\theta)  p(\h)}{q(\h|\x,\phi)} q(\h|\x,\phi)d\h \approx
\]
\[
  \approx \frac 1K \sum_{k=1}^K \frac{p(\x|\h^{(k)},\theta)  p(\h^{(k)})}{q(\h^{(k)}|\x,\phi)},
\]
where $K$ samples $\h^{(k)}$ are obtained from the inference net 
$\h^{(k)} \sim q(\h|\x,\phi)$
via reparametrization trick. 
For the target hypothesis the situation is more complicated. To make computation feasible we assume that $\x_1$ and $\x_2$ are conditionally independent given $\h$:
\[
  P(\x_1,\x_2\mid \theta) = \int p(\x_1, \x_2\mid\h,\theta)  p(\h) d\h = 
\]
\begin{equation}
  = \int p(\x_1\mid\h,\theta)  p(\x_2\mid\h,\theta)  p(\h) d\h.
\label{eq10}  
\end{equation}
Since such an assumption is specific for training the conventional Gaussian PLDA analyzer model~\cite{PLDA}, it is natural for testing PLDA model as well~\cite{Brummer}. However it is not the case for VAE, where training vectors are fed into the model without speaker labels, in fully unsupervised manner. However, our experiments (see Section~\ref{sec3}) demonstrated that this assumption is highly reasonable, so VAE performs a {speaker embedding} which is discussed in~\ref{subsec3-1}. With using this assumption we can use the importance sampling once again to compute marginal likelihood:
\begin{equation}
  P(\x_1, \x_2 \mid \theta) \approx
  \frac 1K \sum_{k=1}^K \frac{p(\x_1|\h^{(k)},\theta)  p(\x_2|\h^{(k)},\theta)  p(\h^{(k)})}{q(\h^{(k)}|\x_2,\phi)}.
 \label{eq11}  
\end{equation}
This expression is asymmetric, because samples are taken from $q(\h^{(k)}|\x_2,\phi)$ when i-vector $\x_2$ is an enrollment one.
However, under the target hypotheses they could be taken from $q(\h^{(k)}|\x_1,\phi)$ as well. Therefore one can use the symmetric LR estimate, where  
$\widetilde{P}(\x_1, \x_2 | \theta) = (P(\x_1, \x_2 | \theta)  + P(\x_2, \x_1 | \theta)/2$
takes both these sampling variants into account. However we found no significant difference between $P(\x_1, \x_2 \mid \theta)$ and $\widetilde{P}(\x_1, \x_2 \mid \theta)$ in our experiments on NIST-2010 (DET-5)~\cite{NIST}. That is why all results shown below were obtained with the use of  $P(\x_1, \x_2 \mid \theta)$ in log-LR estimate, i.e. on the assumption of feeding enrollment vector into inference net.


\subsection{$\beta$-VAE}
In the recent paper~\cite{Higgins} on $\beta$-VAE the empirical deviation from the exact lower bound was used:
\[
   \mathcal{L}(\x) =   \mathbb{E}_{q(\h\mid\x)}\left[ p(\x | \h) \right] 
                                - \beta   D_{KL}\left[q(\h|\x) \mid\mid p(\h)\right].
\]  
The KL-divergence term in~(\ref{eq8}) can be treated as a natural regularizer (which follows from the variational Bayes) for the lower bound. It was observed in~\cite{Higgins} that if VAE is trained with $\beta>1$ (i.e. with high penalty on the likelihood term) then it can better disentangle factors than with the theoretical value $\beta=1$. 
In the speaker recognition domain the factors are, for example, eigenvoices in PLDA model. In Section~\ref{sec3} we demonstrate the results of our experiments on investigating $\beta$-VAE in both ``hard'' ($\beta>1$) an ``soft'' ($0<\beta<1$) modes.


\section{Experiments and discussion}
\label{sec3}

All our experiments were carried out for two homogeneous cellular corpora, namely NIST and RusTelecom. Train part for the NIST corpus consists of 17486 sessions from 1763 male speakers taken from NIST 1998-2008. Tests were carried out on the male part of NIST 2010 (DET-5 extended protocol)~\cite{NIST}. Train part of RusTelecom database consists of 116678 sessions from 6508 male speakers and test part consists of 235 male speakers. The details of the extraction of 400-dimensional i-vectors for the NIST corpus with using English ASR DNN and 600-dimensional i-vectors for the RusTelecom corpus with using Russian ASR DNN are described in~\cite{Pekhovsky}. 
For the correct comparison of VAE and PLDA the latter should have diagonal covariances for both noise and posterior components. Here we moved from PLDA with latent variable to a simple diagonal two-covariance model~\cite{Pekhovsky}.
All input vectors for both PLDA and VAE experiments were centered, whitened and length-normalized in both training and testing. Hereinafter we denote a {whitening matrix} as $U$. We used full matrix $U_{full}$ and diagonal matrix $U_{diag}$ for the full-covariance PLDA and diagonal PLDA respectively.

\subsection{Speaker Embedding VAE}
The fact that we selected i-vectors features and thus limited the effectiveness of VAE by that of PLDA is very convenient. By carrying out extensive comparison of VAE and PLDA for $\beta=1$ (see Tables~\ref{tab3} and~\ref{tab4}) we can obtain two conclusions at once: 
\begin{itemize}
\item the correctness of LLR-score~(\ref{eq9}),
\item the confirmation of the assumption~(\ref{eq10}).
\end{itemize}
The second conclusion states that VAE performs {speaker embedding} in space of latent variable $\h$. In other words, similar to PLDA,  for the target hypothesis VAE is able to sample $\x_1$ and $\x_2$ from the likelihood $p(\x|\h, \theta)$ conditioned on $\h$ of a single speaker.

\subsection{Exploring $\beta$-VAE in low-dimensional space}
\label{subsec3-1}
The second effect was found during $\beta$-VAE experiments, when we explored the ``soft'' training mode ($0<\beta<1$). 
Carrying out the experiments on synthetic data we found that when $0<\beta<1$ diagonal VAE model starts to behave like full-covariance (in posterior) VAE model being able to capture the observed training data from Gaussian clusters with non-diagonal covariance. In order to investigate this property in real-life speaker verification task we selected 11119 files of 660 male speakers having at least 10 sessions. We used PCA projections of 400-dimensional i-vectors in order to operate with a wide range of VAE's number of parameters under comparatively small training dataset. 
\begin{figure}[!ht]
  \centering
  \includegraphics[width=\linewidth]{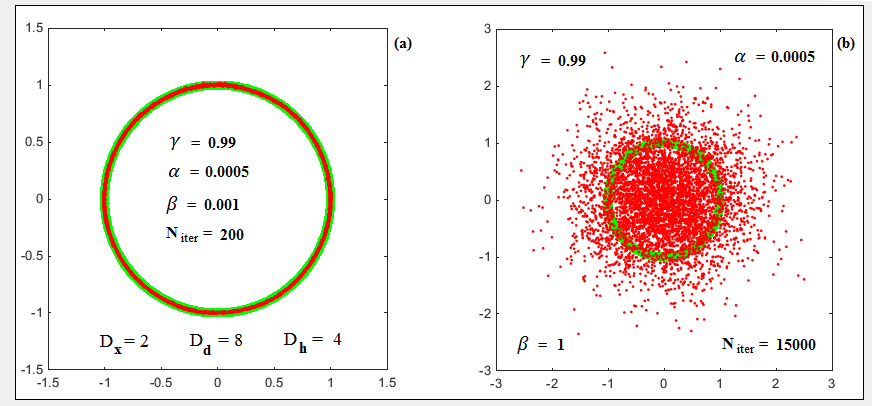}
  \caption{Learning VAE on 11119 NIST files (PCA=2), better viewed in color}
  \label{fig2}
\end{figure}

Figure~\ref{fig2} shows two modes of $\beta$-VAE training for PCA=2. In Figure~\ref{fig2}a the obvious capture (red points) of 660 speakers training data (green points) is observed for the weak regularization mode ($\beta<1$). This differs from the standard mode ($\beta=1$) shown in~\ref{fig2}b. We found that the necessary condition for such a behavior is not only weak regularization but also the sufficient number of neurons in both stochastic and deterministic layers. For instance we unable to achieve this capture for the configuration $\{D_x=2; D_d=2; D_h=2\}$, the minimum configuration required is $\{D_x=2; D_d=4; D_h=4\}$. Our explanation is as follows. Since the expressiveness of the VAE model depends a lot on a posterior power then when increasing a number of posterior diagonal covariance elements to $D_h=4$ we can expect that the capabilities of the diagonal covariance VAE will be strengthened up to those of the full-covariance VAE. Anyway we can assert that making hidden layer $\h$ wider than deterministic ones is necessary for such behavior.

In order to find out if this effect is only a result of overfitting or not and if it may be useful in speaker verification, we carried out a number of verification experiments at PCA=10 for the different VAE configurations. To avoid a strong overfitting we performed a verification tests on the rest 6367 files out of total 17486 in parallel to training. And we stopped training when EER and minDCF metrics computed on this development set started to degrade. Then we tested the obtained VAE model on the male part of the NIST 2010 (DET-5 extended protocol)~\cite{NIST} using $K=100$ for estimating LR-score~(\ref{eq9}). The results are shown in Table~\ref{tab1}.
\begin{table}[!ht]
\setlength\arrayrulewidth{1pt}
	\caption{\label{tab1} {\it VAE experiments on NIST-2010 (DET-5). $PCA=10$, $K=100$ }}
        \centerline{
            \begin{tabular}{|c|c|c|c|c|}
            \hline
            \rowcolor{Orange}
            Dimension 	& \multicolumn{2}{c|}{VAE ($\beta=1$)} & \multicolumn{2}{c|}{VAE ($\beta=10^{-3}$)} \\
            \rowcolor{Orange}
            \cline{2-5}
           \rowcolor{Orange}
    		$D_x$--$D_d$--$D_h$   &  EER   &  DCF   &    EER  &   DCF \\
            \hline
			10--10--5             &  \textbf{9.89}  &  \textbf{0.994}  &  24.15  &  1.000 \\
            \hline
			10--20--10            &  9.86  &  0.996  &  10.56  &  0.997 \\
            \hline
			10--20--20            &  9.83  &  0.998  &  9.90   &  0.999 \\
            \hline
			10--20--40            &  9.88  &  0.994  &  \textbf{9.78}   &  \textbf{0.995} \\
            \hline
			10--20--100           &  9.95  &  0.997  &  \textbf{9.73}   &  \textbf{0.997} \\
            \hline
			10--20--200           &  9.88  &  0.998  &  9.91   &  0.997 \\
            \hline
			10--100--100          &  9.85  &  0.997  &  10.03  &  0.998 \\
            \hline
            \rowcolor{Orange}
                        & \multicolumn{2}{c|}{PLDA (diag-cov)}& \multicolumn{2}{c|}{PLDA (full-cov)} \\
            \rowcolor{Orange}
            \hline
 			                      &  10.02  &  0.998  &  9.07   &  0.997 \\
            \hline
	        \end{tabular}
    }
\end{table}
It can be seen from Table~\ref{tab1} that, contrary to out prior expectations, in both $\beta$ modes VAE is able to exceed the plateau of diagonal PLDA with respect to EER and minDCF. For the ``soft'' $\beta$-VAE there is a conspicuous extremum on the stochastic layer sizes between $D_h=40$ and $D_h=100$. It is not the case for the standard VAE which provides good results starting from the minimal configuration $\{D_x=10; D_d=10; D_h=5\}$ and right up to the maximal number of parameters which is reasonable to use when training on our small training set of 11119 files. 

We carried out such experiments for several values of PCA dimensions and in all cases we observed the same above behavior for two $\beta$-VAE modes. 
``Soft'' $\beta$-VAE is better than standard one and they both are superior to the diagonal PLDA up to the dimension PCA=15 inclusive. However, starting from PCA=20 the standard VAE becomes worse than diagonal PLDA with respect to EER, though comparable to it with respect to minDCF.
One can observe this behavior up to maximal PCA dimensions (limited by i-vector dimension). In these cases only minimal configurations like one shown in the first line of Table~\ref{tab1} can be trained because of the small training set size.


\subsection{$\beta$-VAE in homogeneous corpus}
Experiments on the original cellular corpus NIST (17486 training i-vectors) without PCA dimensionality reduction, i.e. for $D_x=400$, represent the extreme case of our above observations for large PCA dimensions.
The results are shown in Table~\ref{tab3}. Here for all $\beta$-VAE modes best results are achieved for the configuration 
$\{D_x=400; D_d=100; D_h=50\}$ and switched off RMS-prop ($\gamma=1$). There were 160 iterations of VAE training for it to saturate. Here we also tested ``hard'' $\beta$-VAE ($\beta=4$) and found that its behavior doesn't differ significantly from 
that of standard VAE ($\beta=1$). 
It is interesting that LLR estimate depends only marginally on a number of samples used for both $\beta=1$ and $\beta=4$.  
It seems, one might expect that VAE performance improves when $K$ increases, however this behavior is observed for only ``soft'' $\beta$-VAE.
\begin{table}[!ht]
\setlength\arrayrulewidth{1pt}
	\caption{\label{tab3} {\it VAE experiments on NIST-2010 (DET-5) }}
        \centerline{
            \begin{tabular}{|c|c|c|c|c|c|c|}
            \hline
            \rowcolor{Orange}
           $K$ & \multicolumn{2}{c|}{VAE ($\beta=1$)} & \multicolumn{2}{c|}{VAE ($\beta=10^{-6}$)}& \multicolumn{2}{c|}{VAE ($\beta=4$)} \\
            \rowcolor{Orange}
            \cline{2-7}
            \rowcolor{Orange}
            &  EER    &   DCF   &    EER  &  DCF    &    EER  &  DCF \\
            \hline
			 1             &  3.22  &  0.420  &  3.98  &  0.497  &  3.25  &  0.424 \\
            \hline
			 10            &  3.23  &  0.417  &  3.60  &  0.451  &  3.22  &  0.418 \\
            \hline
			100             &  3.23   &  0.417  &  3.31   &  0.449  &  3.24  &  0.418  \\
            \hline
            \rowcolor{Orange}
             	 & \multicolumn{2}{c|}{PLDA (diag)} & \multicolumn{2}{c|}{PLDA (full)}& \multicolumn{2}{c|}{} \\
            \rowcolor{Orange}
            \hline
			     &  3.13   &  0.433  &  1.67   &  0.347  & & \\
            \hline
	        \end{tabular}
    }
\end{table}
As our experiments show such a behavior of LLR estimate is determined not by a difference in $\beta$ modes. The main factor here is a tightness of the lower bound  during the training. Indeed the situation similar to that for the ``soft'' $\beta$-VAE in Table~\ref{tab3} is observed if the tested model is underfit.

\begin{table}[!ht]
\setlength\arrayrulewidth{1pt}
	\caption{\label{tab4} {\it VAE experiments on RusTelecom }}
        \centerline{
            \begin{tabular}{|c|c|c|c|c|}
            \hline
            \rowcolor{Orange}
            $K$ & \multicolumn{2}{c|}{VAE ($\beta=1$)} & \multicolumn{2}{c|}{VAE ($\beta=10^{-5}$)} \\
            \rowcolor{Orange}
            \cline{2-5}
	        \rowcolor{Orange}
            &  EER   &   DCF   &    EER  &  DCF \\
            \hline
			1             &  2.53  &  0.509  &  2.66  &  0.513 \\
            \hline
			10            &  2.53  &  0.511  &  2.58  &  0.510 \\
            \hline
			100            &  2.53  &  0.510  &  2.54   &  0.510 \\
            \hline
            \rowcolor{Orange}
                        & \multicolumn{2}{c|}{PLDA (diag-cov)}& \multicolumn{2}{c|}{PLDA (full-cov)} \\
            \rowcolor{Orange}
            \hline
 			                      &  2.52  &  0.512  &  1.63   &  0.644 \\
            \hline
	        \end{tabular}
    }
\end{table}
In order to improve conditions for training ``soft'' $\beta$-VAE we moved to a larger training corpus of Russian speech, RusTelecom database~\cite{Pekhovsky}. In these experiments the optimal configuration was $\{D_x=600; D_d=400; D_h=200\}$ and RMS-prop was switched off. The learning rate was piecewise-constant starting from $1e-6$ and decreasing once in the middle of training.
The number of iterations was 220. The results shown in Table~\ref{tab4} demonstrate that we managed to slightly improve the ``soft'' $\beta$-VAE situation. However we should have even larger training datasets to achieve the results comparable to those of full-covariance PLDA with ``soft'' $\beta$-VAE.

\section{Conclusions}
The VAE-based speaker verification system in i-vector space is proposed. The LLR estimate for VAE is developed which demonstrates high effectiveness in all experiments with VAE. We showed that VAE performs a speaker embedding during training and thus, contrary to PLDA, can be trained in a fully unsupervised manner on large unlabeled datasets. We found that $\beta$-VAE can be trained in a ``soft'' mode which results in that its properties are close to those of full-covariance VAE model. Last, we demonstrated that in i-vectors space the effectiveness of standard diagonal VAE tends to the plateau corresponding to diagonal PLDA. Therefore we conclude that application of VAE in other features space is of interest.

\bibliographystyle{IEEEtran}

\bibliography{mybib}

\end{document}